\documentclass[%
 reprint,
 amsmath,amssymb,
 aps,
floatfix,
]{revtex4-2}

\usepackage{xcolor}
\usepackage{graphicx}
\usepackage{dcolumn}
\usepackage{bm}
\usepackage{mathtools,amssymb,lipsum}

\usepackage[unicode=true,
 bookmarks=true,bookmarksnumbered=true,bookmarksopen=true,bookmarksopenlevel=1,
 breaklinks=false,pdfborder={0 0 0},pdfborderstyle={},backref=false,colorlinks=false]
 {hyperref}
\hypersetup{pdftitle={Chiral terahertz lasing with Berry curvature dipoles},
 pdfauthor={Kasra Rouhi},
 pdfpagelayout=OneColumn, pdfnewwindow=true, pdfstartview=XYZ, plainpages=false}

\usepackage{fancyhdr}

\begin{document}

\title{Chiral terahertz lasing with Berry curvature dipoles}

\author{Amin Hakimi $^1$}
\email{amin.hakimi@uci.edu}

\author{Kasra Rouhi $^1$}
\email{kasra.rouhi@uci.edu}

\author{Tatiana G. Rappoport $^2$}
\email{tgrappoport@fisica.uminho.pt}

\author{M\'ario G. Silveirinha  $^3$}
\email{mario.silveirinha@tecnico.ulisboa.pt}

\author{Filippo Capolino $^1$}
\email{f.capolino@uci.edu}

\affiliation{$^1$Department of Electrical Engineering and Computer Science, University of California, Irvine, CA 92697, USA
\\
$^2$Centro de Física das Universidade do Minho e do Porto (CF-UM-UP), Universidade do Minho, Portugal
\\
$^3$Instituto Superior Técnico and Instituto de Telecomunicações, University of Lisbon, Avenida Rovisco Pais 1, Lisboa, 1049-001 Portugal
}

\begin{abstract}
Materials with Berry curvature dipoles (BDs) support a non-Hermitian electro-optic (EO) effect that is investigated here for lasing at terahertz (THz) frequencies. Such a system is here conceived as a stack of low-symmetry 2D materials.
We show that a cavity made of such a material supports a nonreciprocal growing mode with elliptical polarization that generates an unstable resonance leading to self-sustained oscillations.
Notably, we demonstrate that the chiral nature of the gain derived from the Berry dipole allows for the manipulation of the laser light's handedness by a simple reversal of the electric field bias.
\end{abstract}

\maketitle

%\section{Introduction}
Several optical and electronic phenomena are intricately connected to the geometry of electronic wavefunctions in solids, which is encoded in the Berry curvature \cite{Xiao2010}. Although generally associated with magnetic materials, the Berry curvature -- which is odd under inversion symmetry -- can be finite in momentum space in acentric non-magnetic crystals \cite{Xiao2010}. Recent proposals suggest that electro-optic (EO) effects in this type of material can induce nonreciprocal optical \emph{gain}, with the gain/dissipative response controlled by the light polarization and propagation direction \cite{lannebere2022nonreciprocal,rappoport2023engineering, Shi2023}. This effect is rooted in the system's Berry curvature dipole (BD) \cite{sodemann2015quantum,Xu2018,Zhang2021}, the first moment of Berry curvature integrated over occupied states.

The BDs play a crucial role in various nonlinear electronic and optical effects. Notably, they enable a second-order nonlinear Hall effect in the absence of a magnetic field \cite{Ma2018,Kang2019,du2021nonlinear} and the rectifying of alternating current \cite{Kumar2021,Min2023}. Materials possessing a finite BD can host a kinetic magnetoelectric effect, where an electrical current generates a net magnetization  \cite{Shalygin2012,Furukawa2017,Calavalle2022}. The BD is associated with various optical phenomena, including the kinetic Faraday effect, where the rotatory power is proportional to the bias electrical current and reverses sign with the bias \cite{Vorobev1979,Tsirkin2018, Konig2019}, and the circular photogalvanic effect, where the photocurrent direction is locked to the helicity of the incoming light \cite{Ivchenko1978,Asnin1978,Deyo2009,Xu2018}.

\begin{figure}
\centering	\includegraphics[width=1\columnwidth]{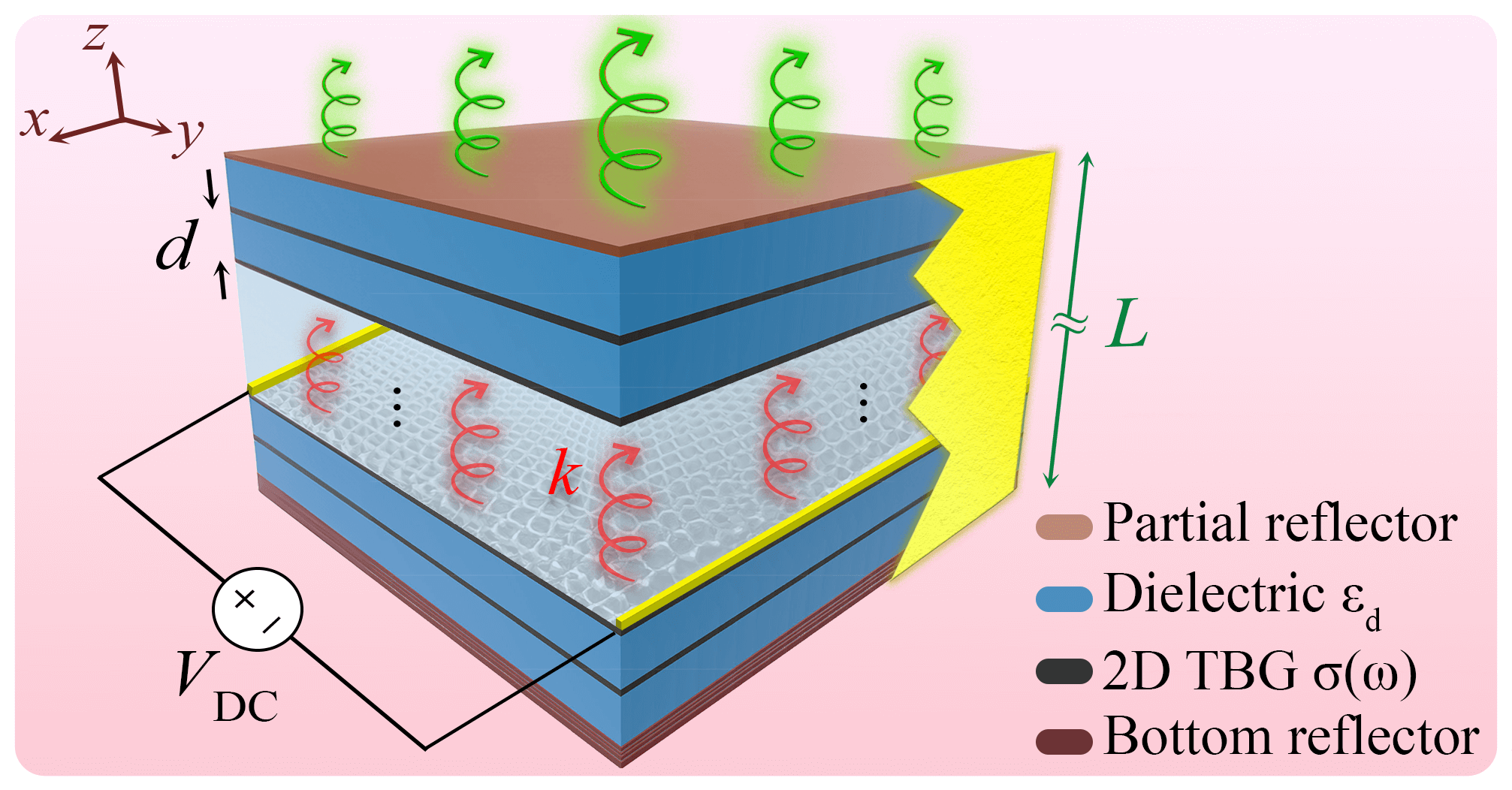}
\caption{Cavity made of a non-Hermitian material. The BDs are in the 2D low-symmetry materials separated by dielectric of thickness $d$. THz radiation through the partial reflector at the top has either right or left handedness. The handedness of the top emission is flipped by reversing the voltage bias.}
\label{fig:MultiLayer3D}
\end{figure}

In this letter, we present a theoretical study that underscores the potential application of the non-Hermitian linear EO effect in BD materials for terahertz (THz) lasing. We propose a mechanism for drawing energy from an appropriately biased BD material into a THz cavity. Notably, our laser exhibits nearly circular polarization, and the handedness of the emitted light is locked to the electric bias sign (Fig. \ref{fig:MultiLayer3D}). Our findings may be relevant for advancing THz and light technologies and offer new insights into novel applications of BD materials. 
Our theory builds on the total conductivity response of a generic low-symmetry 2D material derived in \cite{rappoport2023engineering},

\begin{equation}
\underline{\boldsymbol{\sigma}}(\omega)=
\frac{\sigma_0}{\gamma-i \omega}\left[\begin{array}{cc}
\omega_\mathrm{F} & \xi \\
0 & \omega_\mathrm{F}
\end{array}\right]-\frac{\sigma_0}{\gamma}\left[\begin{array}{cc}
0 & -\xi \\
\xi & 0
\end{array}\right].
\label{eq:sigma}
\end{equation}
Here, $\sigma_0=2 e^2 / h$ is the conductance quantum, $\gamma=1 / \tau$ is the scattering rate, and $\omega_\mathrm{F}=E_\mathrm{F} / \hbar$ where $E_\mathrm{F}$ is the Fermi level. The parameter $\xi$ has units of $\mathrm{s}^{-1}$ and rules the linear EO response; it is proportional to the applied static electric field ($E_0$) and to the BD of the material ($D_{\rm B}$): $\xi  = \pi e D_{\rm B}{E_0}/\hbar$. 

Without an electric bias ($\xi=0$), the longitudinal optical conductivity is dominated by Drude's contribution $\sigma^{(1)}(\omega)=\sigma_\mathrm{D}\left(E_\mathrm{F}\right) /$ $(\gamma-i \omega)$, with $\sigma_\mathrm{D}\left(E_\mathrm{F}\right)=\sigma_0 \omega_\mathrm{F}$. For simplicity, we neglect the anisotropic response of the 2D material so that $\sigma_{x x}^{(1)}(\omega) =\sigma_{y y}^{(1)}(\omega)$.
%The anisotropy has little impact on the non-Hermitian EO effect.
%, which justifies our approximation.

\begin{figure*}
\centering
\includegraphics[width=1\textwidth]{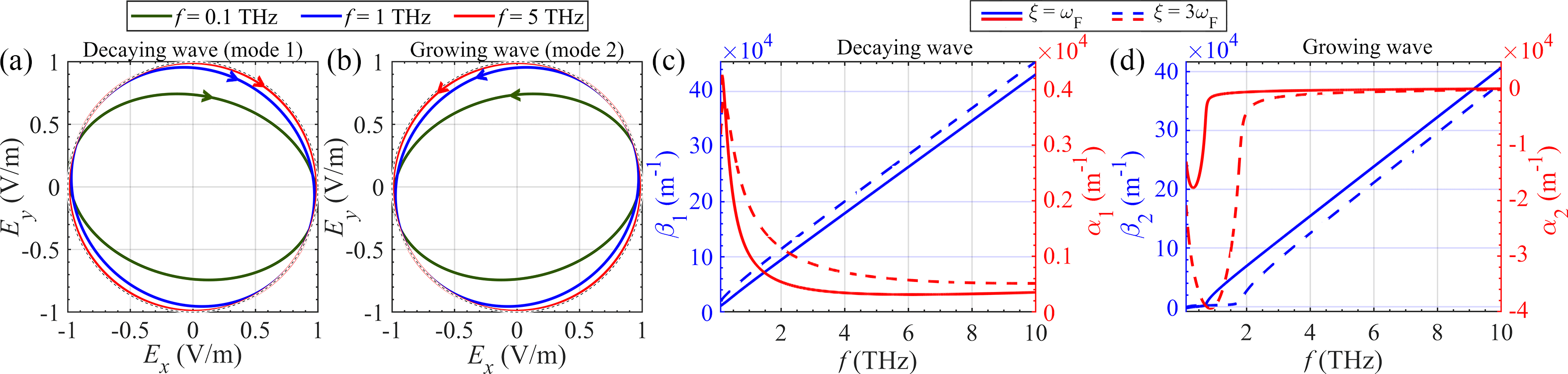}
\caption{Left: Polarization curves and handedness for the (a) decaying and (b) growing waves at three different frequencies. Right: The real ($\beta_i$) and imaginary ($\alpha_i$) parts of the complex wavenumbers: (c) $k_1$ (decaying wave) and (d) $k_2$ (growing wave) with gain parameter $\xi = \omega_\mathrm{F}$  (solid lines) and $\xi = 3\omega_\mathrm{F}$  (dashed lines) by varying frequency.}
\label{fig:PolStates&k123}
\end{figure*}

With a static bias, as in Fig. \ref{fig:MultiLayer3D}, the optical response gets two additional components proportional to $\xi$: (i) a frequency-dependent term that can generate optical gain, in the first matrix in Eq. (\ref{eq:sigma}), and (ii) a frequency-independent part that models a gyrotropic conservative response (second matrix in Eq. (\ref{eq:sigma})). The optical gain originates from non-Hermitian light-matter interactions mediated by the BD. 
A large BD is desirable to maximize these effects. In 2D, the BD components can be expressed in terms of contributions from states near the Fermi surface, $D_{\rm B}^{a}=\int\frac{d^2k}{(2\pi)^2}\Omega_\mathbf{k}\frac{\partial f^0_\mathbf{k}}{\partial k_a}$, where $\Omega_\mathbf{k}$ is the Berry curvature and $f^0_\mathbf{k}$ is the Fermi-Dirac distribution \cite{sodemann2015quantum}.
The BD depends on the product of the Berry curvature and the distribution function derivative. Thus, the most promising candidates for large BDs are systems with narrow gaps, like those achieved through nanopatterning 2D materials \cite{ho2021hall} or in twisted bilayers \cite{he2021giant, pantaleon2021tunable,Duan2022,Sinha2022,Kang2023,Huang2023}. These systems concentrate band velocity and Berry curvature near highly localized Dirac cones, enhancing the BD \cite{he2021giant}. Very large BDs have also been observed in oxide interfaces \cite{Lesne2023,Mercaldo2023}.

Here, we assume that the non-Hermitian material is composed of a stack of 2D low-symmetry materials with large BD which are modeled by Dirac-like Hamiltonians. As an example, we consider twisted bilayer graphene (TBG) layers. 
%\section{Propagation in a material with transverse Berry dipoles}
The layers are biased by a D.C. electric field directed along the $y$ direction that induces a drift current. The BDs provide the mechanism to transfer energy from drifting electrons to the THz field \cite{rappoport2023engineering}. According to what is discussed below, these non-Hermitian interactions are controlled by the handedness of the THz field. Each dielectric spacer has a subwavelength thickness $d$ and relative permittivity $\varepsilon_\mathrm{d}=\varepsilon^\prime_\mathrm{d}+i\varepsilon^{\prime\prime}_\mathrm{d}$. The 2D material layers are isolated from their neighbors, so that their electronic responses are independent. 

Within the stack of 2D materials the electromagnetic field satisfies 
$\nabla \times \textit{\textbf{H}}=-i \omega \varepsilon_0 \varepsilon_\mathrm{d} \textit{\textbf{E}}+\textit{\textbf{J}}$, where $\textit{\textbf{J}}=\delta(z) \underline{\boldsymbol{\sigma}} \cdot \textit{\textbf{E}}_\mathrm{t}$ is the surface current density along the sheet and $\textit{\textbf{E}}_\mathrm{t}$ is the transverse component of the electric field. For simplicity, we characterize wave propagation in the multilayer structure using an effective medium approximation (EMA) that models the system as a homogeneous medium.

Following the analysis of \cite{othman2013graphene} for a stack of isotropic graphene layers, we spatially average the electric and displacement fields over one period in the $z$ direction, leading to an equivalent homogeneous anisotropic medium with effective relative permittivity $\underline{\boldsymbol{\varepsilon}} =\boldsymbol{\underline{\varepsilon}}_\mathrm{t} +\varepsilon_\mathrm{z} \hat{\mathbf{z}} \hat{\mathbf{z}}$.
As the 2D material is infinitesimally thin compared to the dielectric spacer thickness, the longitudinal permittivity is simply 
$\varepsilon_{z}=\varepsilon_{\mathrm{d}}$, whereas the transverse effective relative permittivity dyad takes the form $
\boldsymbol{\underline{\varepsilon}}_\mathrm{t}=\varepsilon_\mathrm{d} \underline{\mathbf{I}} +i \underline{\boldsymbol{\sigma}}/(\omega \varepsilon_0 d)$, where $\underline{\mathbf{I}}$  is the identity dyad.

We focus on wave propagation along the $z$ direction of homogenized (i.e., spatially averaged) fields $\boldsymbol{\mathrm{E}} \propto e^{i k z}$, with $\boldsymbol{\mathrm{E}}$ confined in the transverse plane. Wave propagation properties are fully determined by the dyad $\boldsymbol{\underline{\varepsilon}}_\mathrm{t}$ that  
%We assume that the material is described by a non-Hermitian relative permittivity tensor $\underline{\boldsymbol{\varepsilon}} =\boldsymbol{\underline{\varepsilon}}_\mathrm{t} +\varepsilon_\mathrm{z} \hat{\mathbf{z}} \hat{\mathbf{z}}$. Since in this paper we focus on wave propagation along the $z$ direction, we will only consider field components in the $x-y$ plane and hence we consider only the relative \emph{transverse} permittivity tensor $\boldsymbol{\underline{\varepsilon}}_\mathrm{t}$.
is explicitly written in matrix form as
\begin{equation}
\boldsymbol{\underline{\varepsilon}}_\mathrm{t}=\left[\begin{array}{cc}
\varepsilon_\mathrm{a} & 
\varepsilon_\mathrm{b}\\
\varepsilon_\mathrm{c} & 
\varepsilon_\mathrm{a}
\end{array}\right],
\end{equation}
where the elements of the non-Hermitian matrix are   $\varepsilon_\mathrm{a}=\varepsilon_\mathrm{d}+i \frac{\omega_0 \omega_\mathrm{F}}{\omega (\gamma-i \omega)}$, $\varepsilon_\mathrm{b}=i\frac{\omega_0}{\omega} \xi\left(\frac{1}{\gamma-i\omega}+\frac{1}{\gamma}\right)$ and $\varepsilon_\mathrm{c}=-i\frac{\omega_0 \xi}{\omega \gamma}$, and $\omega_0=\frac{\sigma_0}{\varepsilon_0 d}$. The eigenvalues of $\boldsymbol{\underline{\varepsilon}}_\mathrm{t}$ are ${\varepsilon}_{1,2}=\varepsilon_\mathrm{a} \pm \sqrt{\varepsilon_\mathrm{b} \varepsilon_\mathrm{c}}$ and the corresponding two polarization eigenvectors are  $\mathbf{E}_{1,2}=\left[\pm \sqrt{\frac{\varepsilon_\mathrm{b}}{\varepsilon_\mathrm{c}}},1\right]^\mathrm{T}$ ($\mathrm{T}$ is the transpose operator). They are rewritten in terms of the TBG parameters as
\begin{subequations}
\begin{equation}
{\varepsilon}_{1,2} = \varepsilon_\mathrm{d} + \frac{\omega_0}{\omega }\left(\frac{-\frac{\omega_\mathrm{F}}{\omega}}{1+i\frac{\gamma}{\omega}}\pm \frac{\xi}{\gamma} \sqrt{\frac{1+2i\frac{\gamma}{\omega}}{1+i\frac{\gamma}{\omega}}}\right),
\label{eq:eigenvalueS}
\end{equation}
\begin{equation}
\boldsymbol{\mathrm{E}}_{1,2}=
\left[\begin{array}{cc}
i\sqrt{\frac{1+i2\gamma/\omega}{1+i\gamma/\omega}} \\
1
\end{array}\right],
\left[\begin{array}{cc}
-i\sqrt{\frac{1+i2\gamma/\omega}{1+i\gamma/\omega}} \\
1
\end{array}\right].
\label{eq:eigenvectorsV}
\end{equation}
\end{subequations}

The eigenvalue $\varepsilon_2$ may have $\Im(\varepsilon_2)<0$ for some frequency and gain parameter $\xi>0$, providing a medium with gain. For large enough frequencies, the dielectric losses $\Im(\varepsilon_\mathrm{d})=\varepsilon^{\prime\prime}_\mathrm{d}$ will dominate the gain provided by the $\xi$ term. While the eigenvalues depend on $\omega_\mathrm{F}$ and $\xi$, the two elliptical polarization eigenvectors depend only on $\gamma$ and $\omega$; in other words, the gain parameter $\xi$ has no effect on the polarization of the eigenstates. Furthermore, the two eigenvectors in Eq. (\ref{eq:eigenvectorsV}) are also eigenvectors of the conductivity matrix in Eq. (\ref{eq:sigma}). 
Therefore, the polarization eigenstates of TBG are the same as those of the homogenized material described by the effective permittivity. Because of the non-Hermiticity of the permittivity, the two polarization eigenstates are not orthogonal. Nevertheless, the permittivity matrix can be diagonalized using this non-orthogonal basis. The two polarization eigenstates are represented in Fig. \ref{fig:PolStates&k123} at three different frequencies, assuming $\gamma = 10^{12}\:\mathrm{s}^{-1}$. They tend to be orthogonal with circular polarization when $ \omega \gg \gamma$.

\begin{figure}
\centering
\includegraphics[width=1\columnwidth]{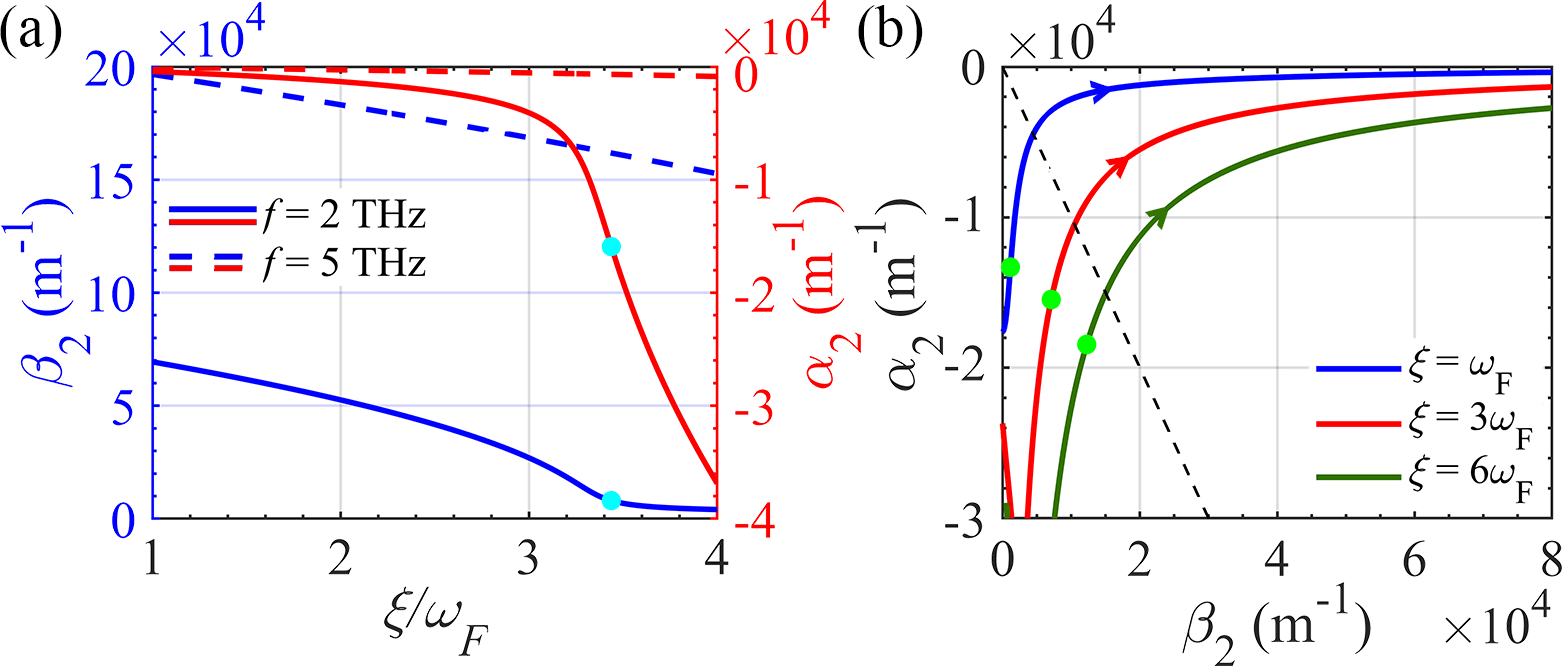}
\caption{(a) Real and the imaginary parts of  $k_2$ (amplifying wave) [Eq. (\ref{eq:k12})] as a function of gain. The light blue dots show when $\xi = \xi_{\mathrm{c}}$. The focus is on the small-gain regime. (b) Complex $k_2$ variation varying frequency, showing both low- and high-gain regimes. 
%The imaginary part versus the real part of the amplifying wave from Eq. (\ref{eq:k12}), while the black dashed line shows the bisector of the fourth quadrant and green dots show when $\omega = \omega_{\mathrm{c}}$.}
The black dashed line is the bisector of the fourth quadrant. Green dots show when $\omega = \omega_{\mathrm{c}}$.}
\label{fig:k2Complexk}
\end{figure}

At high frequencies (i.e., when $ \omega / \gamma \gg 1$ and $\omega /\gamma \gg \omega_\mathrm{F}/\xi$), the eigenvalues are simplified as ${\varepsilon}_{1,2} \approx \varepsilon_\mathrm{d} \pm \frac{\omega_0\xi}{\omega \gamma}$. Since the term $ \frac{\omega_0 |\xi|}{\omega \gamma}$ can be larger than $\varepsilon_\mathrm{d}$,  $\Re(\varepsilon_{1,2})$ can be either positive or negative. Furthermore, for sufficiently large frequencies both $\Re(\varepsilon_{1,2})$ tend to be positive. 
%A negative $\Re(\varepsilon_2)$ results in a strongly growing wave, whereas a positive $\Re(\varepsilon_2)$ will lead to the small-gain regime, as demonstrated later on. 

Because we consider wave propagating along the $z$ direction, the wave equation in the homogenized anisotropic medium is $k_0^2 \,\boldsymbol{\underline{\varepsilon}}_\mathrm{t}  \cdot\boldsymbol{\mathrm{E}} -k^2 \boldsymbol{\mathrm{E}} = 0$, where $k_0 = \omega\sqrt{\mu_0\varepsilon_0}$ is the free space wavenumber (see Supplementary Material). 
It is immediate to observe that the eigenvalues $k^2$ are expressed in terms of the eigenvalues of  $\boldsymbol{\underline{\varepsilon}}_\mathrm{t}$ as  $k^2_{1,2}=k_0^2 {\varepsilon}_{1,2}$, leading to $k_i= k_0 \sqrt{\varepsilon_\mathrm{a} \pm \sqrt{\varepsilon_\mathrm{b} \varepsilon_\mathrm{c}}}$, with $i=1,2$. 
As both $k$ and $-k$ are solutions, in the following we focus only on the two solutions with a positive  $\beta_i=\Re(k_i)$, with $\alpha_i=\Im(k_i)$ representing either the attenuation ($i=1$) or amplification ($i=2$) constant of each mode. The non-orthogonal polarization states of the two electromagnetic modes $i=1,2$ correspond to the two eigenvectors of the non-Hermitian matrix  $\boldsymbol{\underline{\varepsilon}}_\mathrm{t}$ given in Eq. (\ref{eq:eigenvectorsV}). The two wavenumbers are expressed in terms of the TBG parameters as
\begin{equation}
k_i= k_0 \sqrt{\varepsilon_\mathrm{d} + \frac{\omega_0}{\omega }\left(\frac{-\frac{\omega_\mathrm{F}}{\omega}}{1+i\frac{\gamma}{\omega}}\pm \frac{\xi}{\gamma} \sqrt{\frac{1+2i\frac{\gamma}{\omega}}{1+i\frac{\gamma}{\omega}}}\right)}.
\label{eq:k12}
\end{equation}

Figure \ref{fig:PolStates&k123} shows the real and imaginary parts of the wavenumbers, for a material made of a stack of TBG  with $\gamma = 10^{12}\:\mathrm{s}^{-1}$, $\omega_\mathrm{F}/(2\pi) = 0.24\:\mathrm{THz}$, and dielectric spacer with $\varepsilon_\mathrm{d}^{\prime} = 4$, $\varepsilon_\mathrm{d}^{\prime\prime} = 5\times10^{-3}$, and $d = 900\:\mathrm{nm}$ when $\xi = \omega_\mathrm{F}$ (solid lines) and $\xi = 3\omega_\mathrm{F}$ (dashed lines) are used.
These values lead to $\omega_0= 9.7\times10^{12}\:\mathrm{rad/s}$. These are the default parameters used in all simulations. The level of optical gain is consistent with the numerical calculation of the BD of TBG, for an electric field on the order of 0.1 V/$\mu$m and BDs of the order of $D$=10 nm \cite{rappoport2023engineering}. However, the same approach is valid for any 2D material modeled by a tilted Dirac Hamiltonian.
As we can see in Fig. \ref{fig:PolStates&k123}(d), the imaginary part of $k_2$ is negative while the real part is positive. Therefore, for a positive $\xi$, mode 2 grows exponentially along the $z$ direction when $\alpha_2<0$, and mode 1 is instead decaying exponentially ($\alpha_1>0$). 

If one inverts the polarization bias $V_{\mathrm{DC}}$, i.e., the direction of the drift current, the parameter $\xi$ changes sign \cite{rappoport2023engineering}, and the polarization $\mathbf{E}_{1}$ is the one subject to gain. Thus, one may control the handedness of the light emitted by the laser simply by flipping the static bias. Thereby, different from conventional gain systems, a BD material with $\xi>0$ provides gain only when the field polarization has a very specific handedness determined by $\boldsymbol{\mathrm{E}}_2$. 
The fact that mode 2 is amplifying, is verified by the expression of the Poynting vector along the $z$ direction, $\mathrm{Re}(S_{z2})= \frac{\beta_2}{2\omega\mu_0}|\boldsymbol{\mathrm{E}}_2|^2>0$, which confirms the positive power flow in the $+z$ direction while growing exponentially. Note that the $-k_2$ wavenumber solution, also associated with the polarization state $\boldsymbol{\mathrm{E}}_{2}$, grows along the $-z$ direction. 
It is remarkable that since mode 2 is polarized as $\boldsymbol{\mathrm{E}}_{2}$ for both $k_2$ and $-k_2$ wavenumbers, the system is nonreciprocal because the two modes with $k_2$ and $-k_2$ have opposite handedness with respect to the propagation direction. Therefore, we have demonstrated that the non-Hermitian EO effect leads to the amplification of mode 2, in both directions and to the attenuation of mode 1, when $\xi>0$. 
This enables the possibility to have a cavity with a convectional instability relying only on mode 2, leading to lasing. By reversing the bias, the lasing action would be determined by the polarization $\boldsymbol{\mathrm{E}}_{1}$, allowing the laser to radiate a desired handedness by just controlling the biasing voltage $V_{\mathrm{DC}}$ direction. Under the assumption that $\omega \gg \gamma$, the wavenumber of the amplifying mode in Eq. (\ref{eq:k12}) is approximated as

\begin{equation}
k_2 \approx k_0 
\sqrt{\varepsilon_\mathrm{d} - \frac{\omega_0 \xi}{\omega  \gamma}  -\frac{\omega_0} {\omega^2 } \left(\omega_\mathrm{F}+ i\frac{\xi}{2}\right)}.
\label{eq:k12approx}
\end{equation}
There are two gain regimes. The first one is the ``small-gain regime" when $\varepsilon^\prime_\mathrm{d}>\frac{\omega_0\xi}{\omega \gamma}$, leading to the wavenumber in Eq. (\ref{eq:k12approx}) to be mainly real, approximated as  

\begin{equation}
\frac{\beta_2}{k_0}\approx \sqrt{\varepsilon^\prime_\mathrm{d} - \frac{\omega_0\xi}{\omega \gamma}}, \qquad
\frac{\alpha_2}{k_0} \approx - \frac{\frac{\omega_0 \xi}{2\omega^2}-\varepsilon^{\prime\prime}_\mathrm{d}}{2\sqrt{\varepsilon^\prime_\mathrm{d} - \frac{\omega_0\xi}{\omega \gamma}}}.
\label{eq:approxk2Case1}
\end{equation}
The sign of $\alpha_2$ is negative as long as $\omega_0 \xi/(2\omega^2)>\varepsilon^{\prime\prime}_\mathrm{d}$, and it becomes positive at large-enough frequencies. Instead, under the ``high-gain regime" $\frac{\omega_0\xi}{\omega \gamma}> \varepsilon^\prime_\mathrm{d}$, the wavenumber in Eq. (\ref{eq:k12approx}) reduces to

\begin{figure}
\centering
\includegraphics[width=0.95\columnwidth]{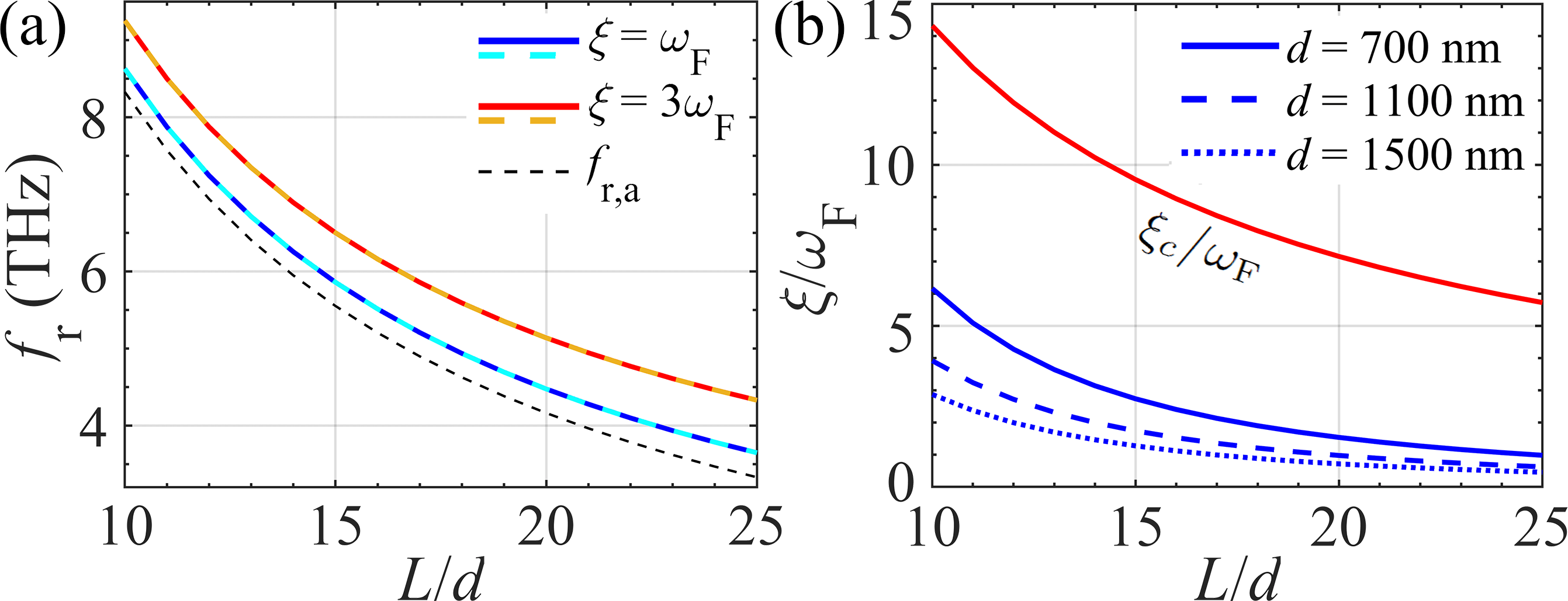}
\caption{(a) Oscillation frequency $f_\mathrm{r}$ versus $L/d$. Solid blue and red curves are solution of Eq. (\ref{eq:OscFreqCond}), and dashed orange and light blue curves are $\Re(\omega/(2\pi))$ from the approximated solution in Eq. (\ref{eq:omega_closedform}). The dashed black curve shows the approximate $f_\mathrm{r,a} \approx c/(2L\sqrt{\varepsilon^{\prime}_\mathrm{d}})$. (b) Normalized threshold gain $\xi_\mathrm{th}/\omega_\mathrm{F}$ for $R=0.99$, varying $L/d$ and the layer thickness $d$. In all three cases, the threshold gain is in the small-gain regime $\xi_\mathrm{th}<\xi_\mathrm{c}$.}
\label{fig:OscFreqThreshold}
\end{figure}

\begin{equation}
\frac{\beta_2}{k_0}\approx  \frac{\frac{\omega_0 \xi}{2\omega^2}-\varepsilon^{\prime\prime}_\mathrm{d}}{2\sqrt{ \frac{\omega_0\xi}{\omega \gamma}-\varepsilon^\prime_\mathrm{d}}} , \qquad
\frac{\alpha_2}{k_0} \approx - \sqrt{\frac{\omega_0\xi}{\omega \gamma}-\varepsilon^{\prime}_\mathrm{d}}.
\label{eq:approxk2Case2}
\end{equation}
This latter high-gain regime condition leads to a large convective amplification rate of mode 2. The two gain regimes show swapped expressions for $\alpha_2$ and $\beta_2$; they are separated by the condition $\varepsilon^\prime_\mathrm{d} = \frac{\omega_0\xi}{\omega \gamma}$, which leads to defining a ``critical gain" given by $\xi_{\mathrm{c}} \approx \omega \frac{ \varepsilon^\prime_\mathrm{d}  \gamma}{\omega_0}$ and a ``critical frequency" as $\omega_{\mathrm{c}} = \frac{\omega_0\xi}{\varepsilon^\prime_\mathrm{d} \gamma}$ that represents the transition between these two gain regimes (see blue and green dots in Figs. \ref{fig:k2Complexk}(a) and (b)). 
%The first case (small-gain regime, when $\xi < \xi_\mathrm{c}$) %with $\alpha_2\approx -\xi \omega_0/(4\omega c \sqrt{\varepsilon^{\prime}_\mathrm{d}}) $ results in slower growth than the second case (high-gain regime).
Establishing a lasing condition within the small-gain regime ( $\xi < \xi_\mathrm{c}$), may also present some advantages due to its larger $\beta_2$.

The volume power density delivered to the EO material is $p_{v} \approx \frac{\omega{\varepsilon}_0}{2} \left[ \varepsilon_{xx}^{\prime\prime}  |\boldsymbol{\mathrm{E}}|^2   +  \omega_0\omega^{-2}\xi\Im \left( E_{x}E_{y}^* \right) \right]$, as shown in the Supplementary Material. Gain occurs when $p_{v}<0$. The formula shows that 
(i) a degree of circular polarization is required to provide a negative value;
(ii) the presence of the BD $D_\mathrm{B}$ is responsible for providing both the gain value $\xi$ and the elliptical polarization of the eigenmodes to make $\Im \left( E_{x}E_{y}^* \right)<0$.

%\section{Cavity with Berry dipoles: Oscillation and threshold gain}
A laser at THz frequencies is conceived by enclosing the material with BDs in a cavity made of two mirrors separated by a distance $L$, as in Fig. \ref{fig:MultiLayer3D}. The growing wave (mode 2) creates self-sustained oscillations, with power emitted from the top surface. We only consider mode 2 propagating within the cavity since mode 1 is attenuating and planar mirrors do not depolarize mode 2. The threshold gain and phase condition for such a self-sustained mechanism is given by
\begin{equation}
Re^{-2\alpha_2 L}e^{i(2\beta_2L-\phi)} = -1.
\label{eq:lasingcond}
\end{equation}
%$Re^{-i\phi}e^{2ik_2L}=-1$, 
where $R$ and $\phi$ are the magnitude and phase of the electric field reflection coefficient at the top partial reflector. For simplicity, we assume that the bottom mirror is a perfect electric conductor providing the reflection coefficient of $-1$. 
%Since the phase propagation constant $\beta_2$ is frequency dependent, the above expression leads to the selection of the first oscillation frequency, given by $2\beta_2 L =\phi + \pi$. 
Furthermore, without loss of generality, we take $\phi=\pi$. Then, the first fundamental oscillation frequency is given by the resonant condition 
\begin{equation}
\beta_2(f)=\pi/L. 
\label{eq:OscFreqCond}
\end{equation}

The cavity resonance $f_\mathrm{r}$ (solution of Eq. (\ref{eq:OscFreqCond})) is provided in Fig. \ref{fig:OscFreqThreshold}(a) (solid curves). Using the approximate solution in Eq. (\ref{eq:k12approx}), a closed-form expression for the first  cavity angular eigenfrequency that satisfies Eq. (\ref{eq:lasingcond}), is

\begin{equation}
\omega \approx \frac{\omega_0 \xi}{2 \varepsilon_d  \gamma} + \sqrt{\left(\frac{\omega_0 \xi}{2 \varepsilon_d \gamma}\right)^2 + \frac{\omega_0}{ \varepsilon_d}\left(\omega_F + i\frac{\xi}{2}\right)+\frac{c^2}{\varepsilon_d}A^2},
\label{eq:omega_closedform}
\end{equation}
where $A = \frac{1}{2L}\left(i\ln{R} + 2\pi\right)$. The real part of this natural frequency is plotted in Fig. \ref{fig:OscFreqThreshold}(a) (dashed blue and orange curves), superimposed on the exact solution obtained from Eq. (\ref{eq:OscFreqCond}) (solid lines).
%Besides the exact frequency solution of Eq. (\ref{eq:OscFreqCond}), or the expression of the approximated one in Eq. (\ref{eq:omega_closedform}), 
The figure also represents $f_\mathrm{r,a} \approx c/(2 L \sqrt{\varepsilon^\prime_\mathrm{d}})$ (dashed black curve), which is obtained by combining Eqs. (\ref{eq:approxk2Case1}) 
and (\ref{eq:OscFreqCond}).

%A physical insight may be inferred by using the approximation of the propagation constant in Eq. (\ref{eq:approxk2Case1}). Indeed, assuming a small-gain regime $ \frac{\omega_0\xi}{\omega \gamma} \ll \varepsilon^\prime_\mathrm{d}$,  one has that $\beta_2\approx k_0 \sqrt{\varepsilon^\prime_\mathrm{d} - \frac{\omega_0\xi}{\omega \gamma}}\approx k_0 \sqrt{\varepsilon^\prime_\mathrm{d}}$. The resonant condition of Eq. (\ref{eq:OscFreqCond}) is then approximated as $f_\mathrm{r,a} \approx c/(2 L \sqrt{\varepsilon^\prime_\mathrm{d}})$, also shown in Fig. \ref{fig:OscFreqThreshold}(a). 

From Eq. (\ref{eq:lasingcond}), a necessary condition for establishing oscillations is that $\alpha_2 < \alpha_\mathrm{th}$, with $\alpha_\mathrm{th} = \frac{1}{2L}\ln(R)$ (i.e., $|\alpha_2| > |\alpha_\mathrm{th}|$, since $\alpha_\mathrm{th}<0$ due to $R<1$).  
An estimate is derived by using the approximation in Eq. (\ref{eq:approxk2Case1}), assuming $\frac{\omega_0\xi}{\omega \gamma} \ll \varepsilon^\prime_\mathrm{d}$, leading to $ \alpha_2 \approx -\xi \omega_0 \beta_2/(4 \varepsilon^\prime_\mathrm{d} \omega^2)$.
This approximation, together with Eq. (\ref{eq:OscFreqCond}), leads to the threshold gain condition
\begin{equation}
\xi_\mathrm{th} \approx -\ln(R) \frac{2\pi c^2}{\omega_0} \frac{1}{L^2}.
\label{eq:lasing_xith}
\end{equation}
The instability of the electromagnetic field within the cavity manifests at the onset of the lasing condition, and it can also be discerned by observing the sign of $\Im(\omega)$. According to Eq. (\ref{eq:omega_closedform}), one has $\Im(\omega)>0$ when $\xi>\xi_\mathrm{th}$. 

%\begin{equation}
%    \xi_\mathrm{th} \approx -\ln(R) \frac{2\pi c^2}{\omega_0} \frac{1}{L^2}.
%\label{eq:lasing_xith}
%\end{equation}

\begin{figure}[t]
\centering
\includegraphics[width=1\columnwidth]{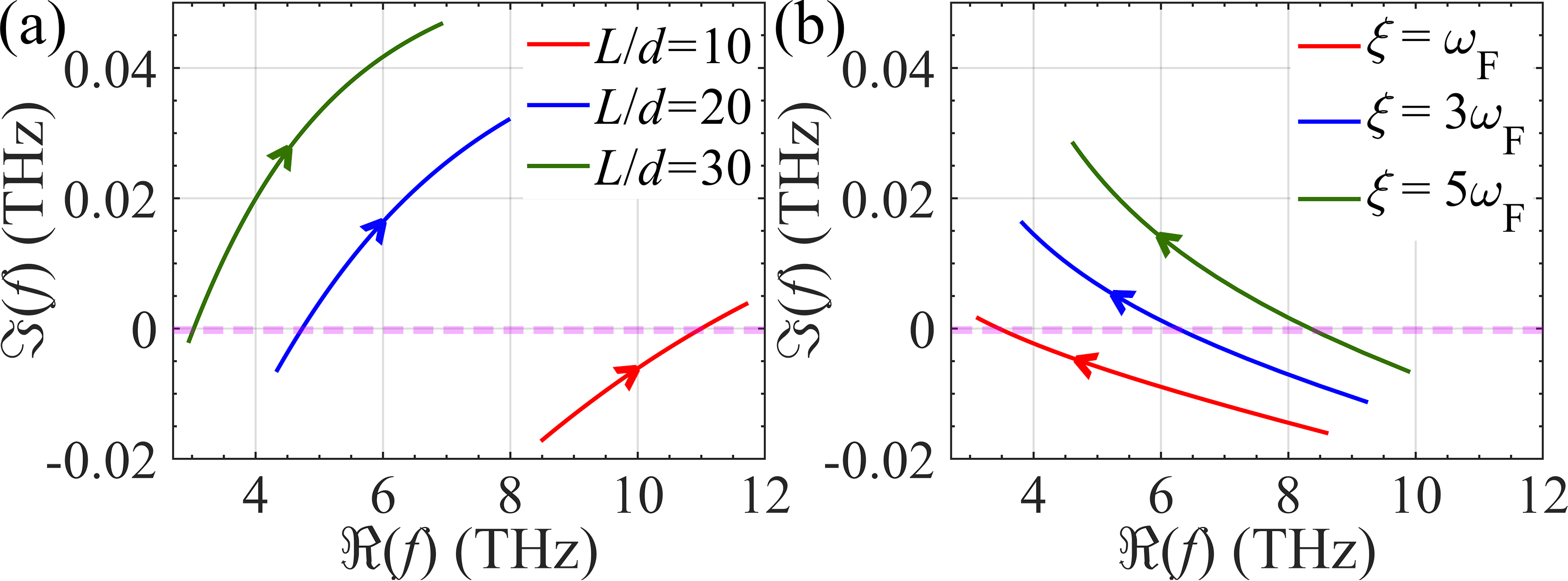}
\caption{Imaginary part versus the real part of the laser natural frequency [Eq. (\ref{eq:omega_closedform})] by varying (a) $\xi$ ($0.5<\xi/\omega_F<10$), and (b) $L$ ($10<L/d<30$). Lasing arises when $\Im(f)>0$. When operating slightly above the threshold, the lasing frequency is roughly $\Re(f)$ at $\Im(f)=0$, which is in the THz region.} 
\label{fig:Complexfreq}
\end{figure}

We determine the scaling laws for the oscillation frequency $f_\mathrm{r} \propto 1/L$ and threshold gain $\xi_\mathrm{th} \propto 1/L^2$ from Eqs. (\ref{eq:OscFreqCond}) and (\ref{eq:lasing_xith}). Furthermore, inserting the estimate of the oscillation frequency $f_\mathrm{r,a}$ into the critical gain expression, leads to the trend $\xi_{\mathrm{c}} \approx \frac{\pi c  \sqrt{\varepsilon^\prime_\mathrm{d}} \gamma}{\omega_0} \frac{1}{L}.$  
Therefore, because of the scaling laws $\xi_\mathrm{th}\propto 1/L^2$ and $\xi_\mathrm{c} \propto 1/L$, by increasing the cavity length it is certainly possible to have a  threshold gain in the small-gain regime. In Fig. \ref{fig:OscFreqThreshold}(b) we show the normalized threshold gain $\xi_\mathrm{th}$ from Eq. (\ref{eq:lasing_xith}) varying  $L/d$; it is possible to operate above threshold leading to lasing under the small-gain regime $\xi<\xi_\mathrm{c}$. The threshold gain is in the range of $\xi_\mathrm{th}/ \omega_\mathrm{F}=1 \textup{--} 4$ for $d>800\:\mathrm{nm}$. 
The results in Fig. \ref{fig:Complexfreq} provide a comprehensive view that it is possible to have gain above threshold at a lasing frequency in the range of 3 THz to 12 THz. 
Any point above the horizontal dashed pink line indicates a cavity signal growing in time (i.e., above threshold). The intersections with the pink dashed line represent the threshold (marginally stable field solution with $\Im(f)=0$), showing also the estimate of the lasing frequency $\Re(f)$. For example, a gain parameter as low as $\xi=\omega_\mathrm{F}$ can induce lasing oscillations at 3.4 THz, for $L/d \approx 27$.

In conclusion, we introduced a new paradigm for THz lasing, leveraging the non-Hermitian linear EO effect in materials with Berry curvature. 
The distinctive feature of our approach lies in the chiral nature of the Berry dipole gain, imparting a well-defined handedness to the laser fields. 
This characteristic can be easily adjusted by inverting the static electric bias. 
Our solution holds great promise for a myriad of applications, offering versatility and adaptability in the realm of THz technology.
\begin{acknowledgements}
M. G. S.  acknowledges  Institution of Engineering and Technology (IET),  the Simons Foundation Award 733700, and  Fundação para a Ciência e a Tecnologia and Instituto de Telecomunicações under Project No. UIDB/50008/2020.
T. G. R.  acknowledges funding from FCT-Portugal through Grant No. CEECIND/07471/2022 
\end{acknowledgements}


%%%%%%%%%%%%%%%%%%%%%%%%%%%%%%%%%%%%%%%%%

%
\onecolumngrid
\newpage
\appendix
\renewcommand{\theequation}{S\arabic{equation}}
\renewcommand{\thefigure}{S\arabic{figure}}
\renewcommand{\bibnumfmt}[1]{[S#1]}
\renewcommand{\citenumfont}[1]{S#1}
\setcounter{page}{1}
\setcounter{equation}{0}
\setcounter{figure}{0}

\begin{center}
{\bf \large{Supplementary material: Chiral terahertz lasing with Berry curvature dipoles}}
\end{center}

\begin{center}
{Amin Hakimi, Kasra Rouhi, Tatiana G. Rappoport, M\'ario G. Silveirinha, Filippo Capolino}
\end{center}

The time notation $\exp(-i\omega t)$ is implicitly used throughout the paper that uses the phasor notation. Furthermore non-italic bold fonts denote the homogenized (i.e., spatially averaged) fields, like $\mathbf{E}$ and $\mathbf{H}$, whereas italic bold symbols, like $\textit{\textbf{E}}$, $\textit{\textbf{H}}$, and $\textit{\textbf{J}}$, denote the non-averaged fields.

\begin{center}
{\bf I. Linear Electro-Optic Response}
\end{center}
Utilizing the semiclassical Boltzmann equation, one can derive the optical (i.e., at THz frequencies) conductivity of a two-dimensional (2D) material subjected to the dynamic field $\textit{\textbf{E}}$ and a static electric bias $\textbf{E}_0$, applied as in Fig. 1 of the paper. In the first-order of both fields, the dynamic current conductivity in the 2D material exhibits an additional electro-optic (EO) contribution induced by $\textbf{E}_0$ as $\textit{\textbf{J}}^{\rm eo}(\omega)=\underline{\boldsymbol{\sigma}}^{\rm eo}(\omega)\cdot \textit{\textbf{E}}$. This contribution arises from anomalous velocity and manifests linearity in both static and dynamic fields. The linearized EO optical conductivity can be decomposed into Hermitian and non-Hermitian contributions $ \underline{\boldsymbol{\sigma}}^{\rm eo}(\omega)= \underline{\boldsymbol{\sigma}}^{\rm eo}_{\rm H}(\omega)+ \underline{\boldsymbol{\sigma}}^{\rm eo}_{\rm NH}(\omega)$:
\begin{equation}
\textit{\textit{\textbf{J}}}^{\rm eo}_{\rm H}=-\frac{e^3\tau}{\hbar^2}(\mathbf{D}_{\rm B}\cdot\textbf{E}_0)(\hat{\mathbf{z}}\times \textit{\textbf{E}})=\underline{\boldsymbol{\sigma}}^{\rm eo}_{\rm H}(\omega)\cdot \textit{\textbf{E}},
\end{equation}

\noindent and 
\begin{equation}
\textit{\textbf{J}}^{\rm eo}_{\rm NH}=-\frac{e^3\tau/\hbar^2}{(1-i\omega\tau)}(\hat{\mathbf{z}}\times \textbf{E}_0)(\mathbf{D}_{\rm B}\cdot\textit{\textbf{E}})=\underline{\boldsymbol{\sigma}}^{\rm eo}_{\rm NH}(\omega)\cdot \textit{\textbf{E}},
\end{equation}
\noindent where $\mathbf{D}_{\rm B}$ is the Berry curvature dipole  (BD) with components 
\begin{equation}
D_{\rm B}^{a}=\int\frac{d^2k}{(2\pi)^2}\Omega^z_\mathbf{k}\frac{\partial f^0_\mathbf{k}}{\partial k_a},
\end{equation}
as explained in the paper and in Ref. \cite{rappoport2023engineeringSupp}. For a Berry curvature dipole $D_{\rm B}$ 
 along  $\hat{\mathbf{y}}$ and an electric field bias $\textbf{E}_0=E_0\hat{\mathbf{y}}$, the two contributions to the EO optical conductivity are given by

\begin{equation}
\underline{\boldsymbol{\sigma}}^{\rm eo}_{\rm H}(\omega)=
-\frac{e^3\tau}{\hbar^2}\left[\begin{array}{cc}
0 & -D_{\rm B}E_0 \\
D_{\rm B}E_0 & 0
\end{array}\right],
\end{equation}

\begin{equation}
\underline{\boldsymbol{\sigma}}^{\rm eo}_{\rm NH}(\omega)=
-\frac{e^3\tau/\hbar^2}{(1-i\omega\tau)}\left[\begin{array}{cc}
0 & -D_{\rm B}E_0 \\
0 & 0
\end{array}\right],
\end{equation}
where the non-Hermitian $\underline{\boldsymbol{\sigma}}^{\rm eo}_{\rm NH}$ is responsible for the process of harvesting energy from the biasing DC field to the optical field.
The diagonal part of the  {\em total}  optical conductivity $\underline{\boldsymbol{\sigma}}$ given by $\sigma_{xx}$ and  $\sigma_{yy}$, originates from the first order terms of the Boltzmann equation while the off-diagonal parts $\sigma_{xy}$ and  $\sigma_{yx}$ contain the electro-optics contributions, leading to the full matrix  

\begin{equation}
\underline{\boldsymbol{\sigma}}(\omega)=
\left[\begin{array}{cc}
\sigma_{xx} & \sigma_{xy} \\
\sigma_{yx} & \sigma_{yy}
\end{array}\right] 
\label{eq:sigmat}
\end{equation}
that is given in Eq. (1) of the paper.  
Taking the total optical conductivity into account, the power transferred from the  THz field to a 2D material is written as
\begin{equation}
p_{\rm{dis}}=\frac{1}{2}\Re \left\{\textit{\textbf{J}} \cdot \textit{\textbf{E}}^{*} \right\} = \frac{1}{2}\Re \left\{ \textit{\textbf{E}}^{\ast} \cdot \underline{\boldsymbol{\sigma}}(\omega) \cdot \textit{\textbf{E}} \right\},
\end{equation}
and it reduces to  
\begin{equation}
p_{\rm{dis}}=\frac{1}{2}\Re \left\{(\sigma_{xx}|E_x|^2+\sigma_{yy}|E_y|^2)+(\sigma_{xy}E_x^{*}E_y+\sigma_{yx}E_xE_y^{*})\right\}.\label{eq:Powerfrom2Dsheet}
\end{equation}

The first term of the equation above contains the diagonal part of the optical conductivity, it is positive and corresponds to dissipation processes. The sign of the second term, which contains the EO response can be negative and when it dominates, there is gain. 

If one considers the optical response of a material modeled by a tilted Dirac Hamiltonian, any anisotropy in the diagonal contribution of the optical conductivity only affects the dissipation and does not influence the gain. As a result, without loss of generality, we can simplify our analysis by considering $\sigma_{yy}(\omega)=\sigma_{xx}(\omega)$. These diagonal components of the optical conductivity are dominated by the Drude's contribution, 
\begin{equation}
{\sigma}_{xx}(\omega)=\frac{\sigma_\mathrm{D}(E_\mathrm{F})}{(\gamma-i\omega)},
\end{equation} 
where $\gamma=1/\tau$ is the scattering rate. 
For the Dirac fermions, $\sigma_{\mathrm{D}}(E_\mathrm{F})=\sigma_0\omega_\mathrm{F}$ where $\sigma_0=2e^2/h$ and $\omega_\mathrm{F}=E_\mathrm{F}/\hbar$, where $h$ is the Planck constant. The Drude contribution is combined with the EO conductivity $\underline{\boldsymbol{\sigma}}^{\rm eo}(\omega)$ leading to the total conductivity in Eq. (1) of the paper. The parameter $\xi  = \pi e D_{\rm B}{E_0}/\hbar$ of the main text is related to the off-diagonal terms of $\underline{\boldsymbol{\sigma}}(\omega)$ and it is responsible for the EO gain of a biased 2D material. In particular, it is the nonconservative piece $\underline{\boldsymbol{\sigma}}^{\rm eo}_{\rm NH}(\omega)$ that is responsible for the non-Hermitian EO effect. 

In the case of twisted graphene bilayers (TBGs), the parameters used here are compatible with the Bistritzer-MacDonald Hamiltonian calculations for strained TBG \cite{rappoport2023engineeringSupp,pantaleon2021tunableSupp,Bistritzer2011Supp}. Still, our approach is valid for any Dirac-like Hamiltonian with finite Berry curvature dipole $D_{\rm B}$.

\begin{center}
{\bf II. Effective Medium Approximation}
\end{center}
In the paper, we analyze wave propagation in a homogeneous anisotropic medium with non-Hermitian relative dyadic permittivity of the form

\begin{figure}
\centering
\includegraphics[width=0.4\textwidth]{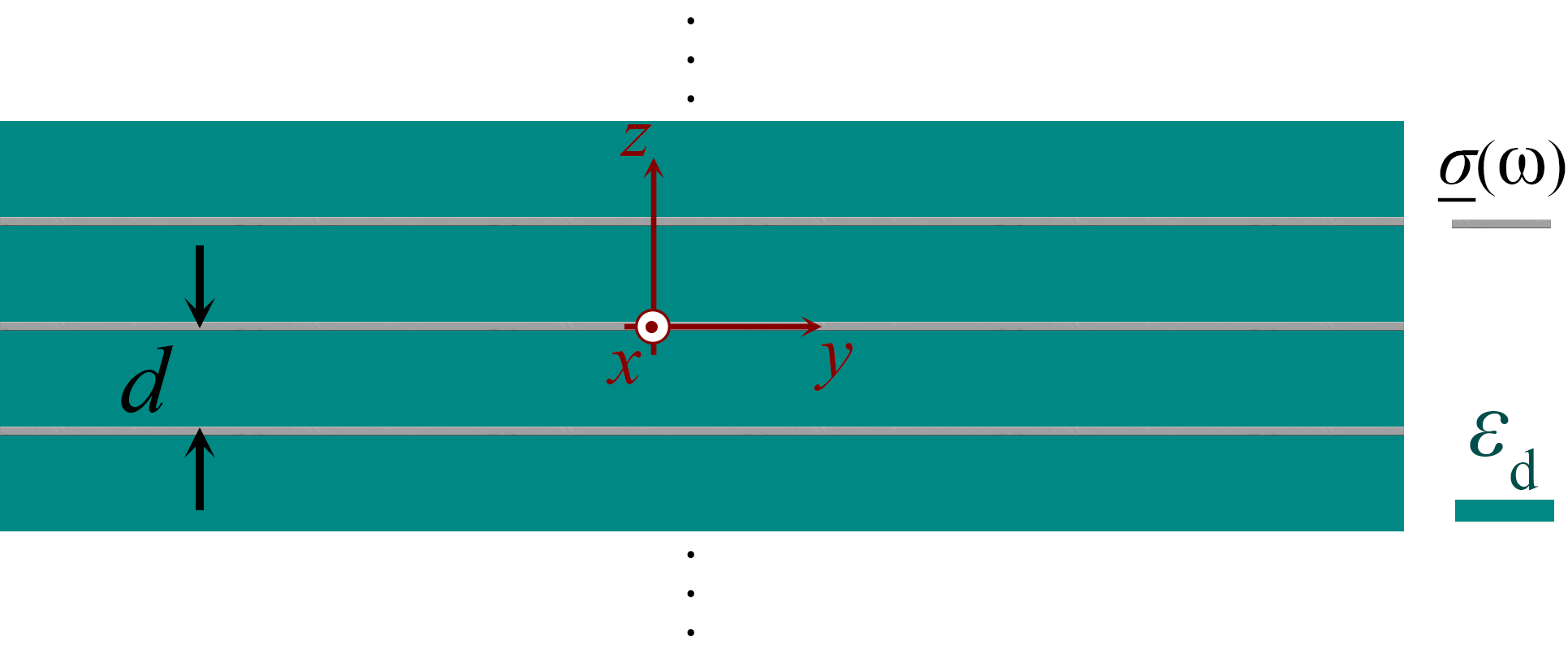}
\caption{Composite multilayer material made by stacking low-symmetry 2D material sheets and dielectric layers. The BDs $D_\mathrm{B}$ are in the low-symmetry 2D sheets that are properly biased by an electric field $E_0$. The multilayered structure is treated as a bulk material with effective bulk permittivity accounting of the electro-optic (EO) effect arising from the BDs. We study wave propagation along the $z$ direction.}
\label{fig:EMA}
\end{figure}

\begin{equation}
\label{eq:effectivePerm}
\underline{\boldsymbol{\varepsilon}}=\underline{\boldsymbol{\varepsilon}}_{\mathrm{t}}+\varepsilon_z \hat{\mathbf{z}} \hat{\mathbf{z}}.
\end{equation}

The information of BDs is embedded in the transverse permittivity dyad $\underline{\boldsymbol{\varepsilon}}_{\mathrm{t}}$. Such anisotropic permittivity is constructed in the paper by considering a multilayered stack of low-symmetry 2D material sheets separated by dielectric spacers of thickness $d$ and relative permittivity $\varepsilon_\mathrm{d}$.  Because of the subwavelength spacing $d$, the multilayer stack is modeled as a homogeneous anisotropic medium by applying the effective medium approximation (EMA) \cite{othman2013grapheneSupp}. It is assumed that low-symmetry 2D material sheets are electronically isolated and that neighboring sheets do not influence their electronic properties. Due to the dielectric spacer's significant thickness with respect to the thickness of 2D material sheets, this assumption is accurate. Each 2D sheet is described by an anisotropic surface conductivity $\underline{\boldsymbol{\sigma}}(\omega)$ as shown in Fig. \ref{fig:EMA}. We consider the Maxwell equation in phasor form, with the implicit time convention $\exp(-i\omega t)$, 

\begin{equation}
\nabla \times \textit{\textbf{H}}=-i \omega \varepsilon_0 \varepsilon_\mathrm{d} \textit{\textbf{E}}+\textit{\textbf{J}} = -i \omega \varepsilon_0\left(\varepsilon_\mathrm{d} \underline{\mathbf{I}}+i \frac{1}{\omega \varepsilon_0}  \underline{\boldsymbol{\sigma}}  \delta(z)\right) \cdot \textit{\textbf{E}},
\end{equation}
where the surface current density $\textit{\textbf{J}}$ [A/m$^2$] in the 2D sheet is written as $\textit{\textbf{J}}=\delta(z) \underline{\boldsymbol{\sigma}} \cdot \textit{\textbf{E}}_\mathrm{t}$ and $\textit{\textbf{E}}_\mathrm{t}$ is the transverse component of the electric field phasor $\textit{\textbf{E}}$, and  $\underline{\mathbf{I}}$ is the unit dyad. Bold italic and bold non-italic fonts here denote non-averaged and averaged fields, respectively. 

By spatially averaging fields over the multilayer period (homogenization process), the above Maxwell equation is rewritten as $\nabla \times \mathbf{H}=-i\omega \varepsilon_0 \underline{\varepsilon}  \cdot  \mathbf{E}$, and the spatially averaged electric and displacement fields are related as $\mathbf{D}=\varepsilon_0 \underline{\varepsilon}  \cdot \mathbf{E}$, where $\underline{\varepsilon}$ is the {\em homogenized} relative permittivity of the multilayer medium. 
The homogenized permittivity is obtained by exploiting the continuity of the $x$ and $y$ components of the electric field, at the boundaries between the 2D sheet and dielectric spacer, and averaging the transverse component of the effective displacement field over a period $d$ along $z$ leads to the \emph{effective relative transverse permittivity} 

\begin{equation}
\boldsymbol{\underline{\varepsilon}}_\mathrm{t}=\varepsilon_\mathrm{d} \underline{\mathbf{I}}+i \frac{\underline{\boldsymbol{\sigma}}(\omega)}{\omega \varepsilon_0 d},
\label{eq:epst}
\end{equation}
that must be considered when dealing with macroscopic averaged fields. Since each 2D sheet is infinitesimally thin compared to the dielectric spacer thickness, we have that the effective longitudinal permittivity is such that $\varepsilon_{z}=\varepsilon_{\mathrm{d}}$, because the $z$-directed electric field would not excite any current in the 2D sheet. 
In summary, the stack of low-symmetry 2D sheets is treated as a material with bulk homogenized effective permittivity as in Eq. (\ref{eq:effectivePerm}), where, using Cartesian coordinates, the transverse relative permittivity is explicitly written in matrix form as
\begin{equation}
\boldsymbol{\underline{\varepsilon}}_\mathrm{t}=\left[\begin{array}{cc}
\varepsilon_\mathrm{a} & 
\varepsilon_\mathrm{b}\\
\varepsilon_\mathrm{c} & 
\varepsilon_\mathrm{a}
\end{array}\right].
\end{equation}
When the 2D low-symmetry layers are made of TBG, the elements of the non-Hermitian matrix are   $\varepsilon_\mathrm{a}=\varepsilon_\mathrm{d}+i \frac{\omega_0 \omega_\mathrm{F}}{\omega (\gamma-i \omega)}$, $\varepsilon_\mathrm{b}=i\frac{\omega_0}{\omega} \xi\left(\frac{1}{\gamma-i\omega}+\frac{1}{\gamma}\right)$ and $\varepsilon_\mathrm{c}=-i\frac{\omega_0 \xi}{\omega \gamma}$, and $\omega_0=\frac{\sigma_0}{\varepsilon_0 d}$.  The electromagnetic properties of the multilayered structure depend on the dyadic conductivity of the low-symmetry 2D material. The matrix has two eigenvalues,  $\varepsilon_{1,2}=\varepsilon_\mathrm{a} \pm \sqrt{\varepsilon_\mathrm{b} \varepsilon_\mathrm{c}}$, as discussed in the paper. When $\Im(\varepsilon_{i})<0$ the material has the capability to provide gain. Furthermore, a negative $\Re(\varepsilon_2)$ results in a strongly growing wave of mode 2, whereas a positive $\Re(\varepsilon_2)$ will lead to the small-gain regime, as demonstrated in the paper. 
This is understood by looking at the properties of the two kinds of modes that propagate in the non-Hermitian medium with BDs, discussed in the next section and in the paper.   

\begin{center}
{\bf III. Wave Propagation in a Homogenized Anisotropic Medium}
\end{center}
Maxwell equations in an anisotropic medium without source are 

\begin{equation}
\left\{\begin{array}{l}
\nabla \times \boldsymbol{\mathrm{E}}=i \omega \mu_0 \boldsymbol{\mathrm{H}} \\
\nabla \times \boldsymbol{\mathrm{H}}=-i \omega \varepsilon_0 \boldsymbol{\underline{\varepsilon}}_\mathrm{t} \cdot\boldsymbol{\mathrm{E}}
\end{array}\right.
\label{eq:Maxwell}
\end{equation}
and electromagnetic wave propagation is governed by

\begin{equation}
\nabla^2 \boldsymbol{\mathrm{E}}-\nabla(\nabla \cdot \boldsymbol{\mathrm{E}})+\omega^2 \mu_0 \varepsilon_0\boldsymbol{\underline{\varepsilon}}_\mathrm{t}  \cdot\boldsymbol{\mathrm{E}}=0.
\end{equation}

In this paper, we only consider waves propagating along the $z$ direction, i.e., $\boldsymbol{\mathrm{E}} \propto e^{i k z}$, with $\boldsymbol{\mathrm{E}}$ confined in the transverse $x-y$ plane. Therefore, the above wave equation simplifies to $\nabla^2 \boldsymbol{\mathrm{E}}+\omega^2 \mu_0 \varepsilon_0\boldsymbol{\underline{\varepsilon}}_\mathrm{t}  \cdot \boldsymbol{\mathrm{E}} = 0$. Since $\nabla^2 \boldsymbol{\mathrm{E}}= -k^2 \boldsymbol{\mathrm{E}}$, we obtain $-k^2  \boldsymbol{\mathrm{E}}+\omega^2 \mu_0 \varepsilon_0\boldsymbol{\underline{\varepsilon}}_\mathrm{t} \cdot \boldsymbol{\mathrm{E}}=0$, which is the same wave equation mentioned in the paper. It may be convenient to rewrite the latter wave equation in matrix form using Cartesian coordinates as

\begin{equation}
\left[\begin{array}{cc}
\varepsilon_\mathrm{a} - (k/k_0)^2 & 
\varepsilon_\mathrm{b}\\
\varepsilon_\mathrm{c} & 
\varepsilon_\mathrm{a} - (k/k_0)^2
\end{array}\right]
\left[\begin{array}{cc}
E_x \\
E_y
\end{array}\right]
=0,
\label{eq:matrixform}
\end{equation}
where $k_0 = \omega\sqrt{\mu_0\varepsilon_0}$ is the free space wavenumber. In order to find the modal wavenumbers $k_i$ ($i=1,2$) the determinant of the matrix in Eq. (\ref{eq:matrixform}) should be zero. Therefore, the wavenumbers and polarization states of the two modes mentioned in the paper are given by

\begin{subequations}
\begin{equation}
k_{i} = k_0\sqrt{\varepsilon_a \pm \sqrt{\varepsilon_b \varepsilon_c}},
\label{eq:eigenvalues}
\end{equation}
\begin{equation}
\boldsymbol{\mathrm{E}}_{1,2}=
\left[\begin{array}{cc}
\sqrt{\frac{\varepsilon_b}{\varepsilon_c}} \\
1
\end{array}\right],
\left[\begin{array}{cc}
-\sqrt{\frac{\varepsilon_b}{\varepsilon_c}} \\
1
\end{array}\right].
\label{eq:eigenvectors}
\end{equation}
\label{eq:eigen}
\end{subequations}
This shows that there are two polarization eigenstates allowed, for a total of four modes, since each eigenstate $\boldsymbol{\mathrm{E}}_{i}$ can propagate either along $+z$ or along $-z$, with $k_i$ or with $-k_i$, respectively. 

As mentioned in the paper, the mode with $k_2=\beta_2+i \alpha_2$ is growing exponentially since $\alpha_2<0$. By itself, this result does not imply amplification because a mode may grow along $+z$ and propagate energy in the $-z$ direction, which means that the mode is actually attenuating. The fact that mode 2 is amplifying is verified by analyzing the Poynting vector of this mode. According to Eqs. (\ref{eq:eigenvalues}) and (\ref{eq:eigenvectors}), the electric field of mode 2 is written as $\mathbf{E} = \mathbf{E}_2e^{ik_2z}$. Using Eq. (\ref{eq:Maxwell}), the magnetic field of mode 2 is written as $\mathbf{H} = \hat{\mathbf{z}}\times\mathbf{E}_2\frac{k_2}{\omega\mu_0}e^{ik_2z}$. Therefore, the Poynting vector of mode 2, $\mathbf{S}_\mathrm{2} = \frac{1}{2} \left(\mathbf{E}_2 \times \mathbf{H}_2^*\right)$, is given by

\begin{equation}
\mathbf{S}_\mathrm{2} = \frac{1}{2}\frac{k_2^*}{\omega\mu_0}|\mathbf{E}_2|^2 \hat{\mathbf{z}}.
\end{equation}
The real part of the Poynting vector $\Re (\mathbf{S}_\mathrm{2}) = \frac{1}{2}\frac{\beta_2}{\omega\mu_0}|\mathbf{E}_2|^2 \hat{\mathbf{z}}$ is along the $z$ direction with a positive sign when $\beta_2>0$, which proves that mode 2 is amplifying (i.e., growing) while propagating energy along $+z$, as discussed in the paper. The same conclusion is derived for the mode polarized as $\mathbf{E}_2$ and propagating along the $-z$ direction, i.e., it propagates energy along $-z$ while it is growing exponentially in the same direction. Therefore, the two modes polarized as $\mathbf{E}_2$, along the positive and negative $z$ directions, are responsible for amplification. Note that these two modes have opposite handedness with respect to their direction of propagation. If instead, the sign of $\xi$ is negative, it is the mode polarized as $\mathbf{E}_1$ and propagating with $k_1$ that will amplify in both the positive and negative $z$ directions. The sign of the voltage bias $V_{\mathrm{DC}}$ controls the sign of $\xi$ and hence the handedness of the mode responsible for the lasing action, leading also to the control of the polarization handedness coming out of the laser.

\begin{figure}
\centering
\includegraphics[width=0.7\textwidth]{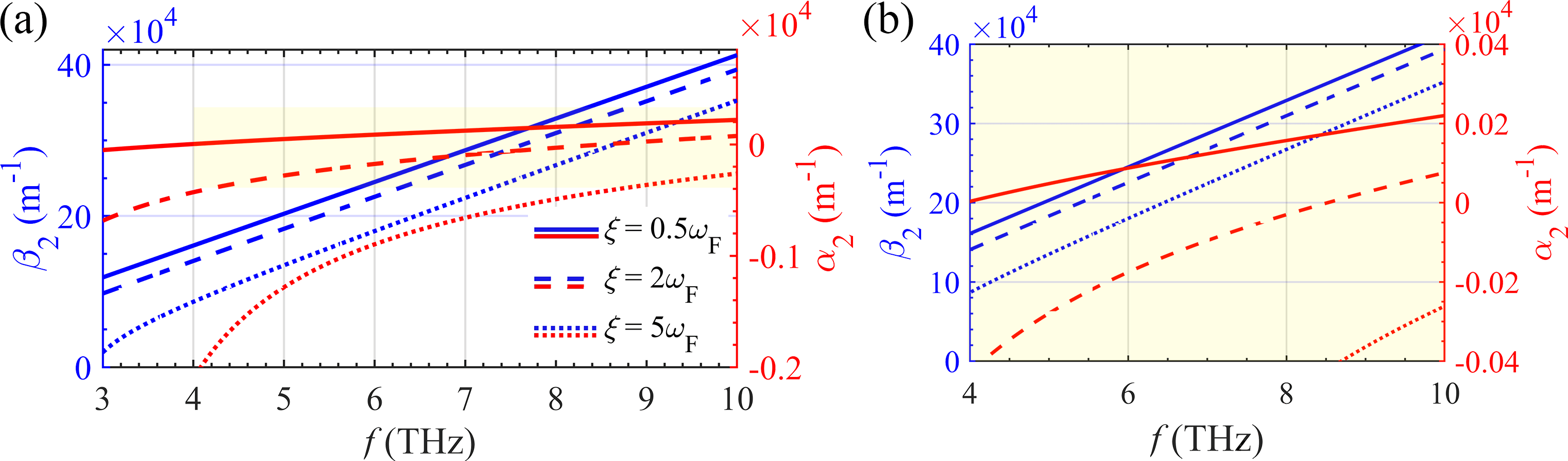}
\caption{The real and imaginary parts of the wavenumber $k_2=\beta_2+i \alpha_2$ (amplifying wave) from Eq. (5) in the paper, by varying frequency, for three values of gain parameters $\xi  = \pi e D_{\rm B}{E_0}/\hbar$. The gain parameter is proportional to the Berry curvature dipole $D_{\rm B}$ and transverse static potential $E_0$. The three cases are all in the small gain regime (i.e., $\xi<\xi_\mathrm{c}$). Amplification occurs when $\alpha_2<0$, which happens when the eigenvalues $\varepsilon_2$ has  $\Im(\varepsilon_2)=\varepsilon_2^{\prime\prime}<0$.}
\label{fig:k2_f}
\end{figure}

Figure \ref{fig:k2_f} shows the real and imaginary parts of the wavenumber of mode 2 (amplifying wave) versus frequency for three values of the gain parameter $\xi$. 
The material is a stack of low-symmetry 2D material sheets with $\gamma = 10^{12}\:\mathrm{s}^{-1}$, $\omega_\mathrm{F}/2\pi = 0.24\:\mathrm{THz}$, and dielectric spacer with $\varepsilon_\mathrm{d}^{\prime} = 4$, $\varepsilon_\mathrm{d}^{\prime\prime} = 5\times10^{-3}$, and $d = 900\:\mathrm{nm}$ (as used in the paper). In this figure, the focus is on the small-gain regime. It is observed that $|\alpha_2| \ll \beta_2$, as expected because of the small-gain regime $\xi<\xi_{\mathrm{c}}$. 
Mode 2 is amplifying because $\alpha_2$ is negative and $\beta_2$ is positive. Furthermore,  $\alpha_2$ is slowly-varying and it is almost constant over this range of frequency. Furthermore, from Fig. (\ref{fig:k2_f}) it is also inferred that mode 2 (amplifying wave) is no longer amplifying after a certain gain cut-off frequency that depends on the gain parameter. As an example, we can see that when $\xi = 2\omega_\mathrm{F}$, the mode is no longer amplifying for $f>8.5\:\mathrm{THz}$. In other words,  $\alpha_2$ switches sign above the gain cut-off frequency and becomes a positive number, which means that the mode is attenuating above the gain cut-off frequency. The reason is due to the losses of dielectric spacer $\varepsilon_d^{\prime\prime}$ and to the scattering rate $\gamma$ of the low-symmetry 2D material.

\subsection{Power transfer from  the EO material to the THz wave}
We analyze the time-averaged power per unit volume delivered by the spatially averaged field to the anisotropic bulk medium with BDs. If the power delivered is negative, it means that the EO material transfers power to the THz wave. This analysis is the bulk analogy to Eq. (\ref{eq:Powerfrom2Dsheet}) that provides the power delivered to a single 2D material. In a homogeneous anisotropic material, the power delivered to the material via electromagnetic wave-matter interaction is given by $P=\int_V p_v dv$, where the time-averaged delivered power per unit volume is given by  \cite{franceschetti1983campiSupp,Felsen1973radiationSupp}

\begin{equation}
p_v=
\frac{1}{2} \Re \left( i \omega {\varepsilon}_0
\boldsymbol{\mathbf{E}} \cdot \boldsymbol{\underline{\varepsilon}}_\mathrm{t}^* \cdot \boldsymbol{\mathbf{E}}^* \right) = \frac{1}{2} \left( \omega {\varepsilon}_0
\boldsymbol{\mathbf{E}} \cdot \boldsymbol{\underline{\varepsilon}}_\mathrm{t}^{\prime\prime^*} \cdot \boldsymbol{\mathbf{E}}^* \right).
\label{eq:power1}
\end{equation}
Here, we decompose the anisotropic permittivity matrix as $\boldsymbol{\underline{\varepsilon}}_\mathrm{t}=\boldsymbol{\underline{\varepsilon}}_\mathrm{t}^{\prime} +i \boldsymbol{\underline{\varepsilon}}_\mathrm{t}^{\prime\prime}$, where $\boldsymbol{\underline{\varepsilon}}_\mathrm{t}^{\prime}$ is the Hermitian part, and $i \boldsymbol{\underline{\varepsilon}}_\mathrm{t}^{\prime\prime}$ is the anti-Hermitian part. Therefore, we have

\begin{equation}
\boldsymbol{\underline{\varepsilon}}_\mathrm{t}^{\prime\prime}= \frac{1}{2i} \left( \boldsymbol{\underline{\varepsilon}}_\mathrm{t}-\boldsymbol{\underline{\varepsilon}}_\mathrm{t}^{\dagger} \right)=  \frac{1}{2i} 
\left[\begin{array}{cc}
\varepsilon_\mathrm{a}-\varepsilon_\mathrm{a}^* & 
\varepsilon_\mathrm{b}-\varepsilon_\mathrm{c}^*\\
\varepsilon_\mathrm{c}-\varepsilon_\mathrm{b^*} & 
\varepsilon_\mathrm{a}-\varepsilon_\mathrm{a}^*
\end{array}\right] =
\left[\begin{array}{cc}
\varepsilon_d^{\prime\prime} + \frac{\omega_0\omega_\mathrm{F}\gamma}{\omega(\omega^2+\gamma^2)} & 
\frac{\omega_0}{2 \omega} \xi\left(\frac{1}{\gamma-i\omega}\right)\\
\frac{\omega_0}{2 \omega} \xi\left(\frac{1}{\gamma+i\omega}\right) & 
\varepsilon_d^{\prime\prime} + \frac{\omega_0\omega_\mathrm{F}\gamma}{\omega(\omega^2+\gamma^2)},
\end{array}\right],
\end{equation}

\noindent where the dagger $\dagger$ denotes the transpose complex conjugation (note that $\boldsymbol{\underline{\varepsilon}}_\mathrm{t}^{\prime\prime}= \boldsymbol{\underline{\varepsilon}}_\mathrm{t}^{\prime\prime\dagger}$ is Hermitian). Using Cartesian coordinates, the dissipated power density is rewritten as

\begin{equation}
p_v  = \frac{\omega{\varepsilon}_0}{2} \varepsilon_{xx}^{\prime\prime*}  \left(
|E_{x}|^2  + |E_{y}|^2 \right)  + \frac{\omega{\varepsilon}_0}{2} \left( \varepsilon_{xy}^{\prime\prime*} E_{x}E_{y}^* + \varepsilon_{yx}^{\prime\prime*} E_{y}E_{x}^*  \right).
\label{eq:power2}
\end{equation}
The term $\varepsilon_{xx}^{\prime\prime}$ is purely real positive so the first term in Eq. (\ref{eq:power2}) is always positive, representing dissipated power density. Furthermore, accounting for the Hermitian property  $\varepsilon_{xy}^{\prime\prime*} =\varepsilon_{yx}^{\prime\prime}$ in the second term in Eq. (\ref{eq:power2}), the delivered power density is rewritten as 

\begin{equation}
p_v  = \frac{\omega{\varepsilon}_0}{2} \varepsilon_{xx}^{\prime\prime}  \left(
|E_{x}|^2  + |E_{y}|^2 \right)  + \omega {\varepsilon}_0 \Re \left( \varepsilon_{xy}^{\prime\prime*}  E_{x}E_{y}^* \right).
\label{eq:power3}
\end{equation}
%\begin{equation}
% \frac{\omega}{2} \left( \varepsilon_{xy}^{\prime\prime}  E_{x}E_{y}^* + \varepsilon_{yx}^{\prime\prime}  E_{y}E_{x}^*  \right) =  \frac{\omega}{2} \left( \varepsilon_{xy}^{\prime\prime}  E_{x}E_{y}^* +  \varepsilon_{xy}^{\prime\prime} ^* E_{y}E_{x}^*  \right) = \omega \Re \left( \varepsilon_{xy}^{\prime\prime}  E_{x}E_{y}^* \right)
%\end{equation}
Note that $\varepsilon_{xy}^{\prime\prime} = \frac{1}{2i}(\varepsilon_{\mathrm{b}}-\varepsilon_{\mathrm{c}}^*)=\frac{\omega_0}{2 \omega} \xi\left(\frac{1}{\gamma-i\omega}\right)$, and for high frequencies $\omega \gg \gamma$, when the polarization eigenstates tend to be circular, one has  $\varepsilon_{xy}^{\prime\prime*} \approx  -i\frac{\omega_0}{2\omega^2} \xi$, and the time-averaged power delivered per unit volume is approximately 

\begin{equation}
p_v  \approx \frac{\omega{\varepsilon}_0}{2} \varepsilon_{xx}^{\prime\prime}  \left(
|E_{x}|^2  + |E_{y}|^2 \right)  +\frac{\omega_0{\varepsilon}_0}{2\omega} \xi \Im \left( E_{x}E_{y}^* \right).
\label{eq:power4}
\end{equation}
From this expression, we observe a few important things: (i) in order to have a negative delivered power density $p_v$, the second term must be negative; (ii) it is clear that a degree of circular polarization is required for the second term to provide a negative number, i.e., a positive power per unit volume delivered to the THz field by the EO material; (iii) it is the presence of the BD $D_\mathrm{B}$  that is responsible for providing both the gain value $\xi$ and the elliptical polarization to make $\Im \left( E_{x}E_{y}^* \right)<0$.   
Applying Eq. (\ref{eq:power4}) to the eigenmode $\boldsymbol{\mathbf{E}}_2$, and assuming high frequency (i.e., $\omega\gg \gamma$), it is clear that the last term is negative, indicating that the polarization eigenmode $\boldsymbol{\mathbf{E}}_2$ is the one associated to a negative power delivered to the material, i.e., a net time-averaged power transferred to the THz wave by the EO material. Vice-versa, when considering the opposite polarization of the eigenstate $\boldsymbol{\mathbf{E}}_1$, the sign of the second term in Eq. (\ref{eq:power4}) is positive, indicating a net power transfer from the THz wave to the material (i.e., attenuation).

When the field takes the form of the eigenmode with polarization $\boldsymbol{\mathbf{E}}_2$, Eq. (\ref{eq:power1}) becomes
 
\begin{equation}
p_v=
\frac{1}{2} \Re \left( i \omega {\varepsilon}_0
\varepsilon_2^* |\boldsymbol{\mathbf{E}}_2|^2 \right) = \frac{\omega}{2} {\varepsilon}_0 
\varepsilon_2^{\prime\prime} |\boldsymbol{\mathbf{E}}_2|^2,
\label{eq:powerMode2}
\end{equation}
where $\varepsilon_2=\varepsilon_2^{\prime}+i\varepsilon_2^{\prime\prime}$ is the second eigenvalue of the matrix $\boldsymbol{\underline{\varepsilon}}_\mathrm{t}$, decomposed in its real and imaginary parts. While this expression is simpler than Eq. (\ref{eq:power4}), it does not show the need for an elliptical polarization to have a negative power density delivered to the EO material, as instead shown explicitly by Eq. (\ref{eq:power4}). 
However, Eq. (\ref{eq:powerMode2}) shows that the volumetric power density transferred by mode 2 to the EO material can be negative when  $\Im(\varepsilon_{2}) = \varepsilon_2^{\prime\prime}<0$. Therefore, mode 2 experiences growth, i.e., $\alpha_2<0$, when $\Im(\varepsilon_{2}) = \varepsilon_2^{\prime\prime}<0$.
Power considerations also imply that when $\varepsilon_2^{\prime\prime}=0$, one has $\alpha_2=0$.
Under the high-frequency approximation $\omega\gg \gamma$, retaining the dominant terms in the Taylor's expansion in Eq. (3a) of the paper, we have that

\begin{equation}
    \Im(\varepsilon_{2}) = \varepsilon_2^{\prime\prime} \approx \varepsilon_d^{\prime\prime} + \frac{\omega_0\omega_\mathrm{F}}{\omega^2}\left(\frac{\gamma}{\omega}-\frac{\xi}{2\omega_\mathrm{F}}\right).
\label{eq:im_eps2}
\end{equation}

\noindent Assuming the loss of the dielectric spacer $\varepsilon_d^{\prime\prime}$ is negligible, mode 2 amplifies while propagating along $z$ when $\xi>2 \omega_\mathrm{F} \gamma/\omega$, and since we assume that $\omega\gg \gamma$, the gain parameter required for amplification is rather small. 
We also observe that the gain threshold value that leads to an amplifying propagating wave (i.e., with $\alpha_2<0$) follows the trend $\xi \propto 1/\omega$, which means that less gain is required at higher frequencies.
However, at very high frequencies, the second term in Eq. (\ref{eq:im_eps2}) becomes small with respect to the losses term $\varepsilon_d^{\prime\prime}$ because of $\omega^2$ at the denominator of the fraction outside the parenthesis.
Therefore, the loss term $\varepsilon_d^{\prime\prime}$ is dominant at very high frequencies, and consequently, mode 2 would not amplify at very high frequencies, even if $\xi>2 \omega_\mathrm{F} \gamma/\omega$. We recall that the gain parameter is related to the BD $D_\mathrm{ B}$ by the formula $\xi = \pi e D_\mathrm{B}{E_0}/\hbar$.

\begin{center}
{\bf IV. Coalescence Parameter and Non-Orthogonality of The Eigenmodes}
\end{center}
The polarization states $\boldsymbol{\mathrm{E}}_i$ of the two electromagnetic modes with $i=1,2$ correspond to the two eigenvectors of the matrix  $\boldsymbol{\underline{\varepsilon}_\mathrm{t}}$ given in Eq. (3b) of the paper. The permittivity matrix $\boldsymbol{\underline{\varepsilon}_\mathrm{t}}$ describing the anisotropic medium is non-Hermitian and the two polarization states are not orthogonal; the angle between them is given by  

\begin{equation}
\cos \theta = \frac{\left|\langle \mathbf{E}_1,\mathbf{E}_2 \rangle\right|}{\lVert \mathbf{E}_1 \rVert\lVert \mathbf{E}_2 \rVert}.
\label{eq:coalescence}
\end{equation}
\begin{figure}[t]
\centering
\includegraphics[width=0.40\textwidth]{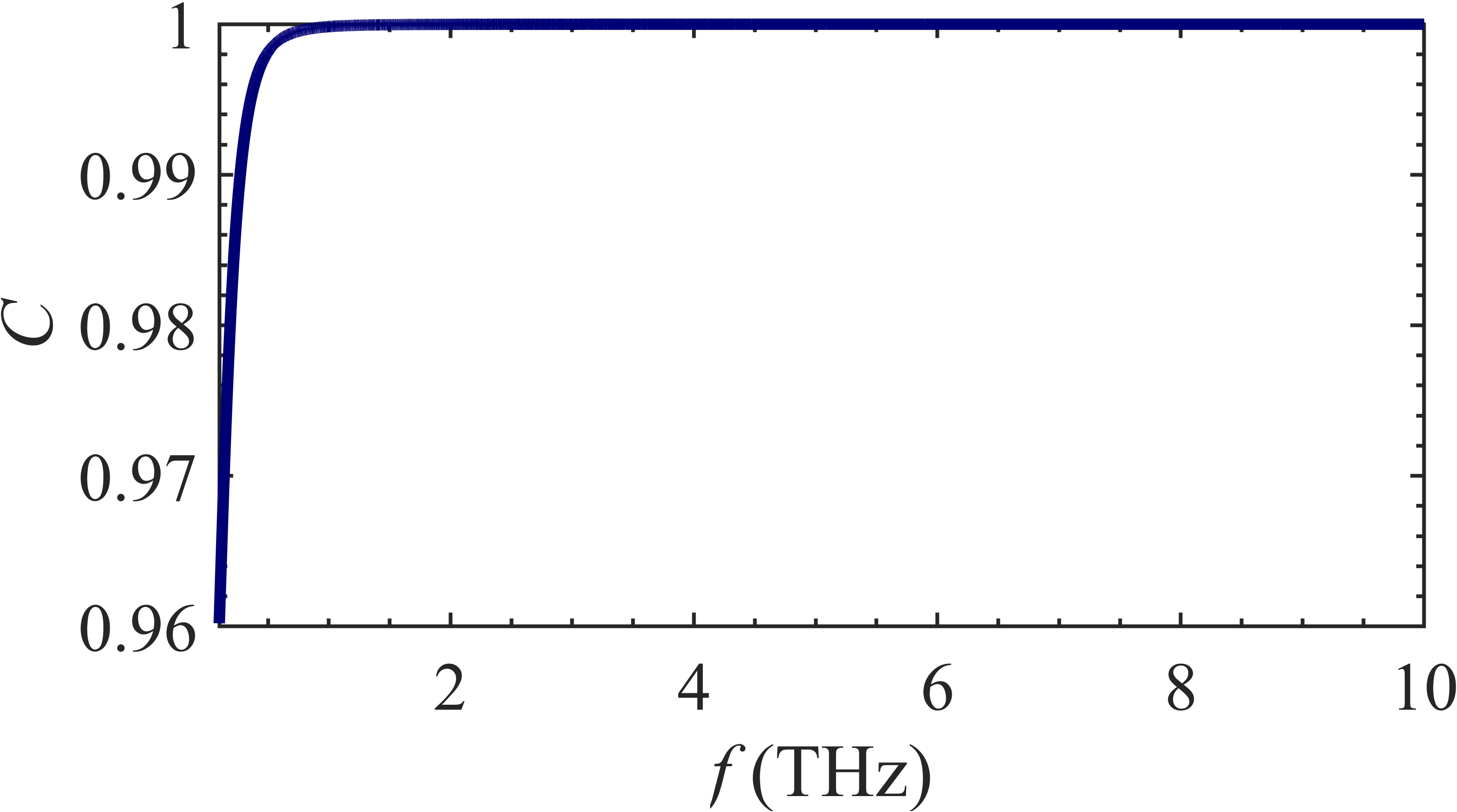}
\caption{Coalescence parameter $C$ versus frequency when $\gamma = 10^{12}\:\mathrm{s}^{-1}$. For higher frequencies, the two eigenvectors (i.e., polarization of the two eigenstates) tend to be orthogonal, having two opposite circular polarizations.}
\label{fig:Coalescence}
\end{figure}
The term $C=\sin \theta$ has been called "coalescence parameter" which can be used to find degenerate states (exceptional points) as already discussed in \cite{abdelshafy2019exceptionalSupp, nada2020frozenSupp}. The coalescence parameter helps to understand the relation between the two eigenvectors. After applying some simplification to Eq. (\ref{eq:coalescence}), the angle expression reduces to

\begin{equation}
\cos \theta = \frac{\sqrt{1+4(\gamma/\omega)^2} -\sqrt{1+(\gamma/\omega)^2}}{\sqrt{1+4(\gamma/\omega)^2}+\sqrt{1+(\gamma/\omega)^2}}.
\end{equation}
The two eigenvectors tend to be parallel when $\theta \rightarrow 0$, or equivalently when $C \rightarrow 0$. In contrast, in our structure, the two eigenvectors tend to be orthogonal when $ \gamma \ll \omega$ since $\cos \theta \rightarrow 0$. Figure \ref{fig:Coalescence} shows that the two polarization eigenstates are neither orthogonal nor parallel in general. However, at very high frequencies where $\gamma \ll \omega$, they tend to be orthogonal and circularly polarized.

\begin{figure}[b]
\centering
\includegraphics[width=0.45\textwidth]{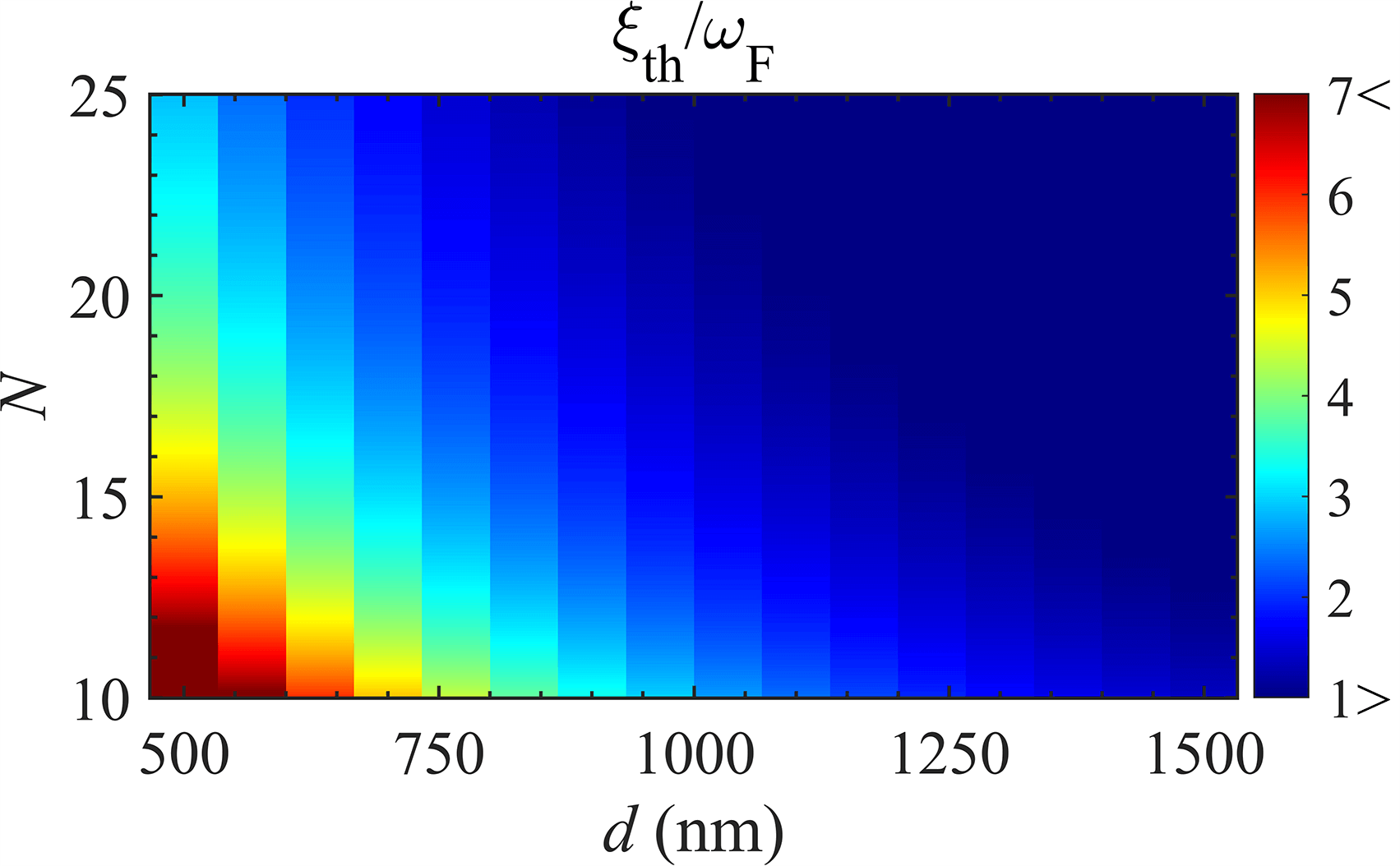}
\caption{Normalized threshold gain $\xi_\mathrm{th}/\omega_\mathrm{F}$ to establish THz lasing in a cavity as in Fig. 1, assuming a reflection coefficient $R=0.99$ from the top partial reflector, varying the number $N$, and thickness of the dielectric spacers $d$.}
\label{fig:LasThreshold_Nd_2D}
\end{figure}

\begin{center}
{\bf V. Threshold Gain Analysis}
\end{center}
The gain parameter threshold for lasing is given in Eq. (11) of the paper. Here we further analyze that equation. Note that in the small-gain regime ($\xi<\xi_{\mathrm{c}}$), the real part $\beta_2$ of the wavenumber $k_2$ is much larger than $\alpha_2$ and therefore it is possible to realize cavities that are not large compared to the free space wavelength since the natural real frequency is given by  $\beta_2=\pi/L$. Furthermore, having $|\alpha_2| \ll \beta_2$ implies small gain per unit length. This condition eliminates the need for large values of gain parameter $\xi$, as would be necessary in the ``large-gain regime''. This implies that the small-gain regime requires a smaller drift current in the 2D low-symmetry material than the large-gain regime.

Figure \ref{fig:LasThreshold_Nd_2D} shows the normalized gain parameter threshold  $\xi_\mathrm{th}/\omega_\mathrm{F}$, by varying the normalized length of the cavity $N=L/d$, and the length of each dielectric spacer $d$. As in the main body of the paper, it is assumed that the field reflection coefficient at the top partial reflector is $R = 0.99$.  From Fig. \ref{fig:LasThreshold_Nd_2D} we observe that the threshold gain is very large for small $N=L/d$ and small $d$. From Eq. (11) in the paper, we observe that $\xi_{\mathrm{th}} \propto 1/L^2= 1/(Nd)^2$. Since it may be difficult to have many discrete layers (large $N$), it may be more efficient to increase $d$ and make $N$ as low as possible to establish lasing with a small threshold. 
For example, Fig. \ref{fig:LasThreshold_Nd_2D} shows that if $d>800\:\mathrm{nm}$, it is possible to make a laser at THz frequencies without having a large number of layers and without high gain. 


%TC:endignore

\end{document}